\documentclass[aps,pre,twocolumn,groupedaddress,nofootinbib]{revtex4-1}
\usepackage[dvips]{graphicx}
\usepackage{amsmath,amsthm,amssymb,color}
\begin{document}

\title{Model of Globally Coupled Duffing Flows}
\author{Tokuzo Shimada}
\author{Takanobu Moriya}
\affiliation{Department of physics, School of Science and Technology, Meiji University \\
1-1-1 Higashimita, Tama, Kawasaki, Kanagawa 214-8571, Japan}
\date{\today}
\begin{abstract}
A Duffing oscillator in a certain parameter range shows period-doubling
that shares the same Feigenbaum ratio with the logistic map, which is an important
issue in the universality in chaos. In this paper a globally coupled lattice of Duffing flows (GCFL), which is a natural extension of the globally coupled logistic map lattice (GCML),
is constructed. It is observed that GCFL inherits various intriguing properties
of GCML and that universality at the level of
elements is thus lifted to that of systems.
Phase diagrams of GCFL are determined, which are essentially the same with those of GCML.
Similar to the two-clustered periodic attractor of GCML, the GCFL two-clustered attractor
exhibits a successive period-doubling with an increase of population imbalance
between the clusters.
A non-trivial distinction between the GCML and GCFL attractors that originates
from the symmetry in the Duffing equation is investigated in detail.\\
\end{abstract}
\maketitle

\section{Introduction}
\label{sec:introduction}
A globally coupled map lattice (GCML) is a system of generally a large number ($N$) of
maps that iteratively evolve  in discrete time ($n$) and interact via their mean field.
One typical example is a homogeneous GCML of logistic maps \cite{kaneko1,kaneko2,kaneko3}:
\begin{equation}
\begin{aligned}
\label{GCML_one_line}
   x_{i}(n+1) &= (1-\varepsilon)f_a(x_{i}(n))+\frac{\varepsilon}{N} \sum_{j = 1}^N f_a(x_{j}(n))   \\
f_a(x)&=1-ax^2,~~~~~~~
~~~~i=1,\cdots,N.
\end{aligned}
\end{equation}

This evolution can be regarded as an iteration of a two-step process.
The first step involves parallel mapping
\begin{equation}
\label{GCML_mapping}
    x^{mid}_i =f \left(x_{i}(n)\right),~i=1,\cdots,N
\end{equation}
where randomness is introduced in the system if the nonlinearity parameter ($a$) of the map is set to be large.
The second step involves interaction with a coupling constant $\varepsilon$ via the mean field:
\begin{align}
\label{GCML_interaction}
   x_{i}(n+1) &= (1-\varepsilon)x^{mid}_{i}+\varepsilon h(n),~i=1,\cdots,N \nonumber\\
   h(n) &\equiv \frac{1}{N} \sum_{i = 1}^N x_i^{mid}.
\end{align}
Here the distribution of maps is contracted to the mean field $h(n)$ at the rate of $1-\varepsilon$
by similarity transformation, and
all the maps are forced to respect their average. Hence, by
this averaging step, coherence is introduced into the system. Note that the relation $\sum_{i = 1}^N x_i(n+1)/N=h(n)$ holds and the average is kept invariant
at the interaction step in this model. This is important for the stability of the model.
This GCML is simple but it shows us how the conflict between randomness and coherence produces
various interesting phases. (Kaneko \cite{kaneko1}).

 However, in a way, GCML is oversimplified. The real physical process
(say, pattern recognition in the brain) occurs in continuous time.
At this point, it is tempting to construct a concise model of the globally coupled flows (GCFL) and
study to what extent it inherits various intriguing properties of GCML. This study aims to search {\it universality in complex systems (GCML and GCFL)}.

 How to choose the flow (as an element of GCFL) that replaces the logistic map (as an
 element of GCML)? According to the Poincar\'e-Bendixon theorem, we must consider a three- or higher-dimensional flow
as an element to exhibit a chaotic behavior at the element level if the flow is autonomous,
or alternatively, one may consider a two-dimensional flow under an external force.
In this paper, we choose, for simplicity, a two-dimensional Duffing
oscillator under an external periodic force
\begin{equation}
\label{duffing_equation}
\frac{dx}{dt}=y, ~\frac{dy}{dt}=-ky-x^3+A\cos t
\end{equation}
with $A=7.5$.\footnote
{This simple parametrization is studied by Yoshisuke Ueda \cite{ueda}, who found
chaos in nonlinear oscillators by analog computer simulation.
In addition, Feigenbaum discusses the Duffing oscillator in this form \cite{feigenbaumreport}.
We have verified that our findings are essentially independent of the choice of the potential shape.
The double-well potential Duffing model also follows the symmetry \eqref{symmetrytransformation},
and similar to our single-well Duffing GCFL, the double-well Duffing GCFL also exhibits
two-clustered attractor under the symmetry \eqref{cluster_pairing} .}
Indeed, it is known that the Duffing flow shares the same period-doubling bifurcation
to chaos with logistic maps, sharing the same Feigenbaum ratio 4.6692016$\cdots$
\cite{cvitanovic,feigenbaumtheory}.
The reason for this universality is as follows:
the two-dimensional flow of this model produces a one-dimensional iterated map
on the Poincar\'e section and, therefore, the bifurcation of this model
is in one-to-one correspondence with the bifurcation of the one-dimensional iterated map.
Besides, all the one-dimensional maps (with a singly peaked function) are governed by a universal function
 at the limiting bifurcation.
The above scenario is crucial in the {\it universality in chaos} \cite{cvitanovic}.
In this paper, we numerically study to what extent the universality at the element level
(the universality between the logistic map and the Duffing oscillator
with respect to the bifurcation to chaos) extends to the similarity
between the systems, i.e., the logistic GCML and the Duffing GCFL.

In fact, the above universality (at the element level)
extends to a certain higher-dimensional flow
and relates it down to a one-dimensional map.
For instance, the three-dimensional Lorenz model \cite{lorenz}
(with $\sigma=10$ and $b=8/3$) has three prominent $r$ bands, in which
it shows period doubling \cite{sparrow}.
The key to this issue is that the Lorenz system is dissipative. Hence,
the effective dimension of the orbit reduces to two in this regime, and on
the Poincar\'e section, it is equivalent to a one-dimensional iterated map
\cite{cvitanovic,feigenbaumtheory}.
Our preliminary analysis of the coupled Lorenz system shows intriguing results:
(1) For strong coupling, a two-clustered attractor is formed just as in the Duffing GCFL.
(2) For weaker coupling, the distribution of flows (at any time
after the transient process) forms a one-dimensional
closed string (single $S^1$ type) in the three-dimensional phase space, and
as we reduce the coupling further, the string bifurcates
at the critical point into two closed strings (two $S^1$ strings linking each other).
This type of topology change is known to occur for a single Lorenz attractor
\cite{jackson}.
We defer the discussion of the Lorenz GCFL to a forthcoming paper.

 Note that the Duffing equation \eqref{duffing_equation} has  symmetry under the transformation
\begin{equation}
\label{symmetrytransformation}
    x \rightarrow -x,~~y \rightarrow -y, ~~t \rightarrow t + \pi
\end{equation},
which allows for pairwise attractors. That is, if $(x_1(t), y_1(t))$ is
a solution of \eqref{duffing_equation}, then
\begin{align}
\label{pairwise-attractors}
  (x_2(t),y_2(t))
  & = (T_{\pi}\otimes P)(x_{1}(t), y_{1}(t)) \nonumber \\
  & \equiv (- x_1(t+\pi), - y_1(t+\pi)),
\end{align}
is also a solution, where the symbols $T_\pi$ and $P$, respectively, denote
the time translation for half period of the external force and the parity transformation
in the two-dimensional phase space.
Depending on the initial value, a Duffing flow may fall in one of the two attractors.
On the other hand, the logistic map has only a single attractor independent of the initial point.
This difference at the element level is reflected in the final structure of the attractors
of coupled models and we will discuss this in detail.

Some results in this paper were reported at AROB 13 \cite{tskktm-arob08}.
This paper presents our completed work with a deepened understanding of GCML-GCFL correspondence (especially relation \eqref{cluster_pairing} and \eqref{system_correspondence}).
On the basis of high statistics simulations, we report below
detailed phase diagrams of the Duffing GCFL and the bifurcation diagram
with respect to the cluster composition ratio $\vartheta$.

 The rest of this paper is organized as follows.
In Sec.\ref{sec:Duffing GCFL}, we first construct a Duffing GCFL along the same lines as GCML.
We put Duffing flows in all-to-all interaction via their mean field.
This interaction acts on the "coordinates" of flows $(x_i(t),y_i(t))$.
At this point, this model differs from most of the other synchronization models
of flows, where elements interact via the time derivatives $(\dot{x}_i(t),\dot{y}_i(t), \cdots)$
\cite{pikovsky, pikovskybook}.
The manner in which the nonlinearity of both GCFL and GCML is matched is
described in detail.
Then, we study the phase structure of the Duffing GCFL by varying the coupling $\varepsilon$
as a free parameter.
We find that the phase diagram of the Duffing GCFL resembles that
of the logistic GCML. The universality at the level
of elements extends to that of coupled systems.
In Sec.~\ref{sec:2_clustered_dynamics}, we focus on
the two-clustered regime of the Duffing GCFL.
Here flows divide themselves into two synchronizing clusters
and the orbits of the two clusters bifurcate
as the population imbalance between the
two clusters increases.
This bifurcation precisely corresponds to the phenomenon
in the two-clustered regime of the logistic GCML discovered by \cite{kaneko1}.
This may also be considered as an extended universality.
However, there is a subtlety owing to the attractor pairing.
We will clarify this issue in Sec.~\ref{sec:2_clustered_dynamics}A.
(Fig.~\ref{Fig:fig6}--\ref{Fig:fig8}).
Only after considering this distinction can we fully understand
both similarity and dissimilarity of nonlinearity reduction in
GCML and GCFL, as discussed in Sec.~\ref{sec:2_clustered_dynamics}B
(Fig.~\ref{Fig:fig9} and \ref{Fig:fig10}).

\section{Duffing GCFL}
\label{sec:Duffing GCFL}

\subsection{Construction of Duffing GCFL}
\label{subsec:Duffing construction}

We construct the Duffing GCFL by replacing the maps in GCML by Duffing flows.
The first step \eqref{GCML_mapping} now becomes
\begin{equation}
\label{GCFL_evolution}
\begin{aligned}
x_i^{mid}(t+\Delta t)&=x_i(t)+y_i(t)\Delta t \\
y_i^{mid}(t+\Delta t)&=y_i(t)+\left[-k y_i(t)-\left(x_i(t)\right)^3+A\cos(t)\right]\Delta t \\
&~~~i=1,\cdots,N
\end{aligned}
\end{equation}
and the second \eqref{GCML_interaction} becomes
\begin{equation}
\label{GCFL_interaction}
    \begin{aligned}
    x_i(t&+\Delta t)=(1-\varepsilon_D)x_i^{mid}(t+\Delta t)+\varepsilon_D h_x , \\
    y_i(t&+\Delta t)=(1-\varepsilon_D)y_i^{mid}(t+\Delta t)+\varepsilon_D h_y ,
    ~~i=1,\cdots,N \\
    h_x &\equiv \frac{1}{N}\sum_{i=1}^{N}x_i^{mid}(t+\Delta t),~~
    h_y \equiv \frac{1}{N}\sum_{i=1}^{N}y_i^{mid}(t+\Delta t).
    \end{aligned}
\end{equation}
The interaction in this step is via the mean field of the system and
the change in the coordinates is directly calculated by a similarity contraction,
similar to that in GCML. Hence, let us designate this model as GCFL.
A similar model was investigated in the earliest
stage of the chaos synchronization study \cite{pecora,kowalski} (Eq.(5) in \cite{kowalski}).
There, the synchronization of the chaotic elements was the main concern,
while in this study, our interest lies in how various synchronizing phases appear
from the originally chaotic flows via nonlinearity reduction due to averaging interaction.
Interesting studies have also been conducted on phase synchronization
(the amplitude may be left unsynchronized) \cite{pikovsky,fns}. Here we study
cluster formation (synchronization in both phase and amplitude).
An extensive account on the synchronization can be viewed in \cite{pikovskybook}.

To investigate the phase structure of the Duffing GCFL, we need some
rough guide of the regions to explore the vast ($k$, $\varepsilon$) parameter space.

\begin{figure}[!bp]
\begin{center}
  \includegraphics[width=80mm,clip]{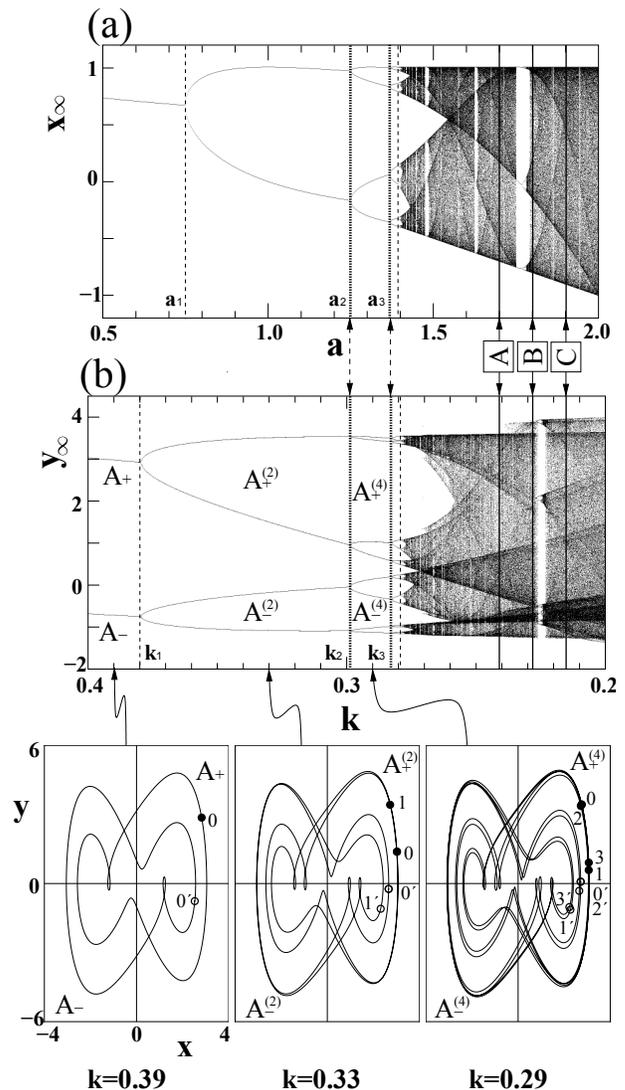}
  \caption{\label{Fig:fig1}
(a) Bifurcation tree of a logistic map. (b) Bifurcation tree of a Duffing flow.
For $k > k_0=0.660$, the Duffing attractor is unique and left-right symmetric (not shown), while
for $k_0 > k > k_1=0.380$, it becomes bivalent (either A${}_{+}$ or A${}_{-}$), each being chiral (see insets).
With a further decrease in $k$, each of A${}_{+}$ or A${}_{-}$ bifurcates. The threshold for $A_{\pm}^{(2^{n-1})} \rightarrow A_{\pm}^{(2^n)}$ is denoted as $k_{n}$.
Arrowed dashed lines connecting (a) and (b) indicate that the second and third bifurcation points are chosen for the matching
of $a$ and $k$. Solid lines show A, B, and C points, which are used for the comparison of GCML and GCFL.}
\end{center}
\end{figure}

Let us first discuss the case of nonlinearity parameters.
To study nonlinearity reduction (i.e., the battle between chaos and coherence),
the nonlinearity parameter of the elements should be set high in the random chaotic regime.
For a logistic map, the threshold to chaos is $a_{th}=1.401$, so let us
choose $a=1.7,1.8, 1.9$ as typical values. As for a Duffing flow,
note that the parameter $k$, the friction constant, is an {\it anti}-nonlinear
parameter. From {\it universality in chaos} \cite{cvitanovic}, in the sense that
two elements in the same universality class have the same bifurcation trees,
a natural choice for comparison is to align the bifurcation tree of a Duffing flow
with that of a logistic map, as shown in Fig.~\ref{Fig:fig1}.

In fact, there is a subtlety in the Duffing flow. At very low nonlinearity (very large friction constant $k$),
the periodic orbit is self-symmetric in the $xy$ plane, and with an increase in nonlinearity,
it changes the topology, splitting into two periodic orbits ($A_{+}$ and $A_{-}$) at ($k \approx 0.66$),
which are mirror symmetric of each other (see insets in Fig.~\ref{Fig:fig1}).
We consider that each of the $A_{\pm}$ corresponds to a fixed point of a logistic map
and align the Duffing tree with the logistic tree
such that the second and third bifurcation points match between the two trees.
\footnote{We are aware that the universality holds at the bifurcation limits,
so in principle, it would be better to consider the matching at the bifurcation points
as high as possible. However, such matching using narrow strips is bound to produce large global errors.
 For a few different alignment choices, we have verified that the main body of our results is unchanged
.}

\begin{figure}[!htbp]
\begin{center}
  \includegraphics[width=80mm,clip]{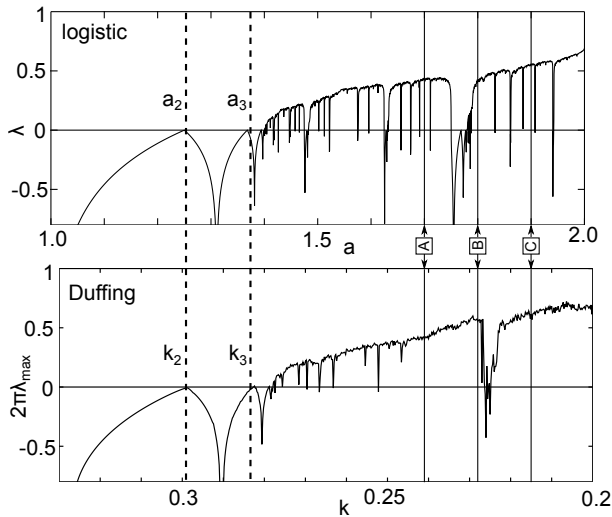}
\caption{\label{Fig:fig2}
Comparison of the element nonlinearity in terms of Lyapunov exponents.
Upper: the exponent of a logistic map
as a function of $a$ in \eqref{GCML_one_line}.
Lower: the maximum exponent of a Duffing flow
as a function of $k$ in \eqref{duffing_equation}.
Dashed lines indicate the second and third bifurcation points, which are chosen for the matching
of $a$ and $k$, and solid lines show A, B, and C points,
which are used for the system comparison
(cf. Fig.~\ref{Fig:fig1}).}
\end{center}
\end{figure}

It is interesting to test how the Lyapunov exponents compare between two models
after matching the bifurcation trees.
The Duffing flow induces a Floquet map at the Poincar\'e shot, which is taken at every
period of the external force. Therefore, we expect a correspondence
\begin{equation}
\label{Lyapunovtest}
 e^{\lambda_\text{flow}\times 2\pi} \approx e^{\lambda_\text{map}}.
\end{equation}
The agreement between the exponents is remarkably good except for the details of windows, as observed in Fig.~\ref{Fig:fig2}.

Up to now, we have been considering matching at the level of the basic elements of two models.
Now, let us consider the coupling $\varepsilon_{\text{D}}$.
In GCML, the similarity contraction by $1-\varepsilon_{\text{map}}$
occurs at each interaction step.
On the other hand, in the Duffing GCFL \eqref{GCFL_evolution},\eqref{GCFL_interaction}, the contraction by $1-\varepsilon_D$
occurs at each $\Delta t$, and $\Delta t$, in turn, has to be chosen sufficiently small
to guarantee finite difference approximation.
(We typically use $\Delta t =10^{-3}$.)
Therefore, the coupling $\varepsilon_D$ must be chosen to be small
enough such that the accumulation of the contraction effect
in GCFL roughly amounts to the contraction in GCML.
That is, very crudely
\begin{align}
\label{epsilon_general_correspondence}
    1 - \varepsilon_{\text{map}} \approx \left( 1 - \varepsilon_{\text{D}}
    \right)^\frac{t_c}{\Delta t},
\end{align}
where $t_c$ is a certain time scale of order one.
If one may simply carry over the correspondence at the level of elements to that
of the systems, one iteration step of a map corresponds to a $2\pi$ evolution
of the flow so that one might assume $t_c=2 \pi$.
However, the systems are evolving while interacting
in nontrivial ways (the crucial point is the existence of a pairing attractor structure in the Duffing flow)
and the above naive expectation is unjustified.

We proceed below without any assumption
on the value of the $\varepsilon$ (and $t_c$).
We regard $\varepsilon$ as a free parameter of the model
and only use \eqref{epsilon_general_correspondence}
as a guide to explore the GCFL phase diagram.
We will find below that in the Duffing GCFL,
phases quite similar to those of the logistic GCML
are realized and the proper choice of $t_c$ is as given in \eqref{system_correspondence}.

\begin{figure}[!bthp]
\begin{center}
  \includegraphics[width=80mm,clip]{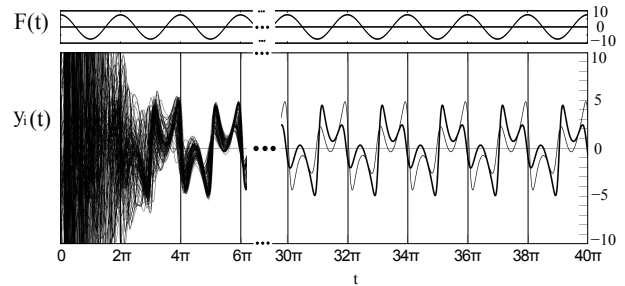}
\caption{\label{Fig:fig3}
A typical run in the Duffing GCFL. $N=100$, $k=0.241$ (case A) and $\varepsilon=1.1 \times 10^{-4}$ ($\Delta t=10^{-3}$).
Upper: periodic external force $F(t)$.
Lower: synchronization of Duffing flows into two periodic clusters.
Only $y$ coordinates are shown.
}
\end{center}
\end{figure}

Now, we are ready to investigate the Duffing GCFL.
Fig.~\ref{Fig:fig3} shows a typical run in the Duffing GCFL with $N=100$. Here the model
($k=0.241$ (case A) and $\varepsilon=1.1 \times 10^{-4}$)
is started at a random initial configuration, and after about 10 cycles of external periodic force,
the flows divide themselves into two synchronizing periodic clusters. This is a typical run for the two-clustered
regime of the Duffing GCFL.  Let us now compare the overall phase structures between the two models.

\subsection{Phase diagrams}
\label{subsec:phase_diagram}

\subsubsection{GCML phase diagram}
\begin{figure}[tbhp]
\begin{center}
  \includegraphics[width=80mm,clip]{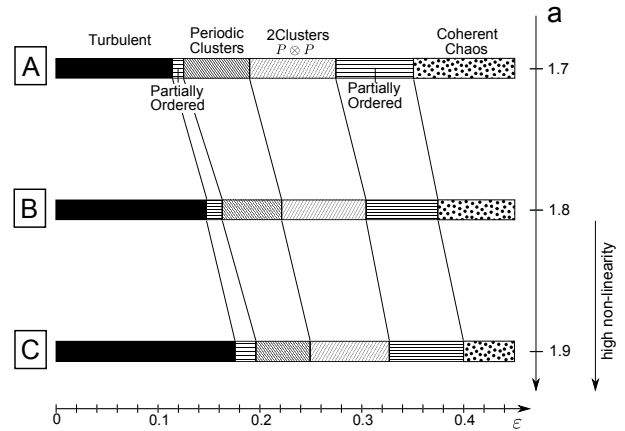}
\caption{\label{Fig:fig4}
GCML phase diagrams at points A, B, and C. (Kaneko \cite{kaneko1}).
With decreasing $\varepsilon$, one observes three outstanding regimes
(coherent chaos, periodic two-clustered, and turbulent regimes, respectively); in between these,
there are periodic-clustered and partially ordered regimes.
With increasing nonlinearity parameter $a$, a larger
coupling $\varepsilon$ is required to maintain the same dynamics.}
\end{center}
\end{figure}
\begin{figure*}[tbhp]
\begin{center}
  \includegraphics[width=140mm,height=100mm,clip]{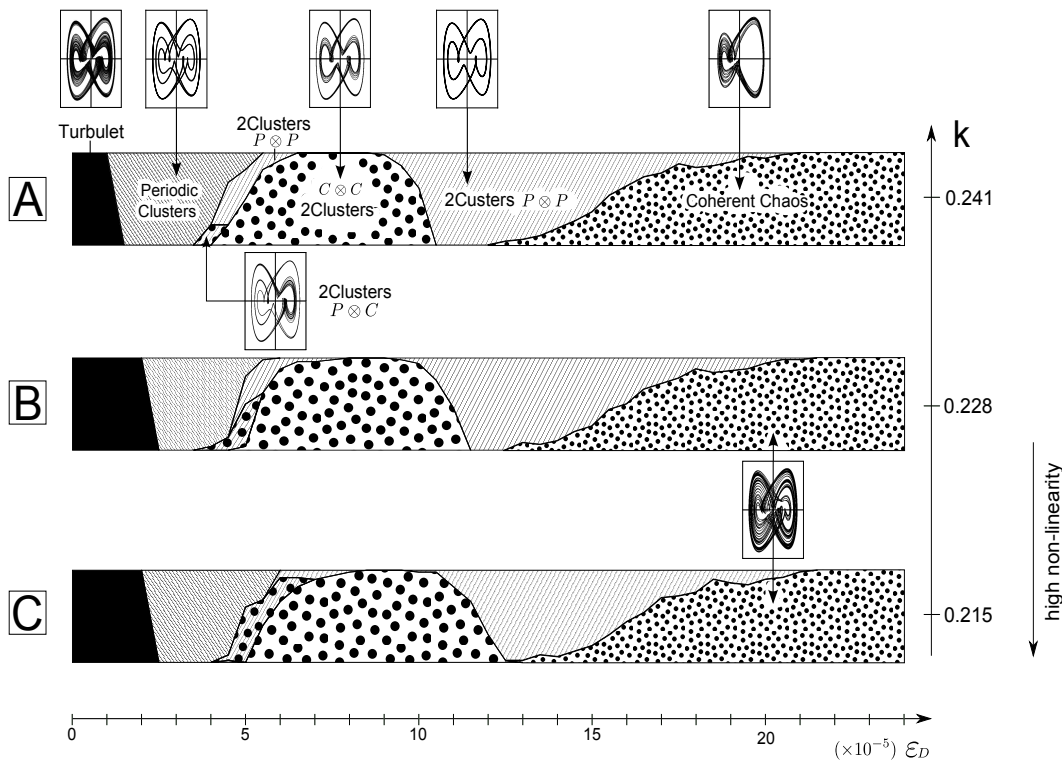}
  \caption{
  \label{Fig:fig5}
Phase diagrams of the Duffing GCFL at
A $(k=0.241)$, B $(k=0.228)$, and C $(k=0.215)$,
corresponding to GCML at A $(a=1.7)$ B $(a=1.8)$, and C $(a=1.9)$, respectively.
(Cf. Figures~\ref{Fig:fig1} and \ref{Fig:fig4}.)
The time step for the finite time approximation is $\Delta t = 10^{-3}$ and
the coupling constant $\varepsilon_D$ is
varied from $0$ to $24\times 10^{-5}$ with increments of $0.5\times 10^{-5}$.
At each $\varepsilon$, 50 random configurations are generated
and the final attractors after $t=500 (2\pi)$ are examined for cluster structures and
orbit types. The insets exhibit the orbits of the flows in the final attractor
in the respective phase. Basin volume ratios of the final attractors are
exhibited by the partition of the bar width.
The attractor of the type {\it coherent-chaos} at the strong coupling regime
changes its topology.
At the low nonlinearity side of the element flow (case A), the attractor is chiral, while
at the high nonlinearity side (B and C), it is self-dual.
With increasing nonlinearity (decreasing $k$), a larger
coupling $\varepsilon$ is required to maintain the same dynamics.
}
\end{center}
\end{figure*}The phase structure of GCML \eqref{GCML_one_line} is explored by Kaneko;
refer to \cite{kaneko1} for a detailed phase diagram in the $(a,\varepsilon)$
plane. To facilitate a comparison with the GCFL phases, we reproduce in
Fig.~\ref{Fig:fig4} the GCML phases at A, B, and C points.
\footnote{We have cross-checked these phase structures.}
There are three outstanding phases:
\begin{itemize}
    \item [(1)]  {\it The coherent chaos state} at large $\varepsilon$.
     The maps are strongly bunched
     together in one cluster ($x_i(n)=X(n),~i=1,\cdots,N$).
     Then, $h(n)=X(n)$ and, therefore, the interaction \eqref{GCML_interaction}
     becomes immaterial.
     All the maps evolve in a bunch just as a single logistic map at the original
     $a$.
    \item [(2)] {\it The two-clustered phase} at intermediate $\varepsilon$.
     The final maps divide into two clusters($\mu=1,2$),
     ($x^\mu_i(n)=X^\mu(n),~i=1,\cdots,N_\mu$), and the two clusters, with populations
     $N_1$ and $N_2$, respectively, oscillate periodically opposite in phase to each other,
     thus keeping the fluctuation of the mean field $h(n)$ small.
     This opposite phase motion is a solution for stability. A remarkable property found by Kaneko \cite{kaneko1} is that
     the population ratio $\vartheta=N_1/(N_1+N_2)$ acts as a new bifurcation parameter.
    \item [(3)] {\it The turbulent phase} at very small $\varepsilon$.
      In general, the number of clusters is proportional to $N$ and
      maps evolve almost independently of the randomness of the original $a$.
    \footnote{
      However, statistical analysis reveals {\it hidden coherence}---
      a long-time-scale correlation \cite{kaneko2}. Furthermore, there occur drastic
      periodic motions of maps in stable clusters at specific $\varepsilon$ values
      tuned with $a$'s---{\it periodicity manifestations such as $p3c3$ and $p3c2$}\cite{tskk-pre}.
      The reflection of these manifest cluster formations dominates all over the turbulent regime of GCML.
      These are due to the foliation of periodic window dynamics of the element
      maps (\ref{subsec:GCML_foliation}). We refer to \cite{shibatakaneko}, where
      the foliation at the small $\varepsilon$ region is caught as a {\it tongue-like structure}.}
\end{itemize}
In addition, in a region marked as {\it periodic clusters},
several clusters ($(2,3)$, $(2,3,4)$, $(3,4), \cdots$) are formed with decreasing $\varepsilon$,
and in {\it partially ordered phases}, the basin volume is partitioned by
the dynamics of two surrounding regions.

\subsubsection{GCFL phase diagram}
Fig.~\ref{Fig:fig5} shows the phase diagrams of the Duffing
GCFL for three choices of the friction coefficient, $k$=$0.241,0.228$, and $0.215$
corresponding to A, B, and C, respectively, in Fig.~\ref{Fig:fig1}.
In the figure, basin volume ratios of the final attractors are
exhibited by the partition of the bar width. For instance,
at $k=0.241$ (point A) and $\varepsilon=(10.4-13.4)\times 10^{-5}$,
the flows almost always (with 10\% exception) fall into
two-clustered attractors (both periodic (P$\otimes$P) as shown in the inset),
while at the same $k$ but at $\varepsilon=15.5$, the basin volume of the
P$\otimes$P attractor and that of the coherent chaotic attractor are almost the same
 (the center of a partially ordered phase).
Comparing Fig.~\ref{Fig:fig5} with Fig.~\ref{Fig:fig4}, we observe that the phase structure of the Duffing GCFL is generally the same as that of
GCML. The three basic phases of GCML, namely (1) the coherent chaos, (2) the two-clustered
phase, and (3) the turbulent phase, are also realized in the Duffing GCFL and in the same
order as coupling strength. Moreover, the tendency that the regimes shift to the higher
coupling with increasing nonlinearity is also the same.
(Note that the lower friction $k$ implies higher nonlinearity.)  The agreement in the basic phase structure
implies that the Duffing GCFL inherits the intriguing properties of the logistic GCML
and that the universality at the element level may be lifted to that
at the level of globally coupled systems.

  Below, we list the nontrivial properties of the Duffing GCFL that are not present in GCML:
\begin{itemize}
    \item [(1)] Symmetry of the Duffing equation under \eqref{symmetrytransformation}
    leads to an attractor consisting of {\it pairwise} clusters.
    The clusters may be called mirror pairs (but with a shift of $\pi$ in time.)
    (On the other hand, in the GCML two-clustered regime, clusters evolve
     in essentially the {\it same} orbit with a shift of one step.)

    \item [(2)] As an intriguing consequence of (1), we observe not only the
    $P\otimes P$ state but also the $C\otimes C$ and $C\otimes P$ states.

    \item [(3)] If the coupling is high enough, the flows are bunched in a
    cluster and the cluster evolves with the original
    high nonlinearity of the elements (coherent chaos). At this point,
    the case is the same as that for GCML. However, the Duffing flow at high nonlinearity
    shows two types of attractors. If the nonlinearity is not so high, the
    chaotic flow is confined in either the left or right polarized orbit (chiral type), while
    at the higher nonlinearity, the chaotic flow extends over the joint
    of the above two orbits (self-dual type). This is reflected in the coherent chaos state.
    (Compare the coherent chaos in $A$ with that in $B$ and $C$).
\end{itemize}

Hereafter, we focus on the interesting two-clustered phase,
which is realized in both GCFL and Duffing GCFL.

\section{Two-Clustered Dynamics of The Duffing GCFL}
\label{sec:2_clustered_dynamics}

\subsection{Attractor $\vartheta$-bifurcation}
\label{subsec:theta bifurcation}

\subsubsection{$\vartheta$-bifurcation in GCML}
In the two-clustered regime, the GCML attractor
consists of two clusters mutually oscillating
opposite in phase. Let $N_+$ be the number of maps in
the {\it positive} cluster (the one consisting of maps
with $x_i(n) > h(n)$ at even iteration step $n$).
Kaneko found that with an increase (or decrease) in the
{\it population ratio} $\vartheta \equiv N_+/(N_++N_-)$,
the cluster attractor repeats successive bifurcations
until the imbalance reaches a certain threshold beyond which
the system goes into chaotic transients \cite{kaneko1,kaneko3}.
Let us call this phenomenon $\vartheta$-bifurcation.
It implies that the cluster attractor is labeled by three parameters
($a$, $\varepsilon$, and $\vartheta$) and it may lead to
a possible new approach of information processing.

\subsubsection{Attractor symmetry and $\vartheta$-bifurcation in GCFL}
 Let us investigate if our Duffing GCFL inherits this intriguing property.
The answer is yes, but, in order to fully understand the
$\vartheta$-bifurcation in GCFL, we have to take account of an attractor symmetry
coming from the symmetry \eqref{symmetrytransformation} of the Duffing equation.

In Fig.~\ref{Fig:fig6}, we show three typical events
in the GCFL two-clustered regime. The central event is basic. It is $\vartheta =0.5$
and is of  period one. If a time translation of $\pi$ is applied on one of the two clusters,
its position becomes exactly mirror symmetric to that of the other cluster.
The other two events are $\vartheta=0.77$ and $\vartheta=0.23$.
The cluster orbits are indeed bifurcated and are of period two ($T=4\pi$).
Let us call the attractors type A and type B if $\vartheta > 0.5$ and
$\vartheta < 0.5$, respectively,
and two attractors as partner attractors if $\vartheta_A + \vartheta_B=1$.
The attractors shown at the right and the left of the basic
attractor, respectively, are thus partner attractors.
Inspecting the motion of clusters in Fig.~\ref{Fig:fig6},
we find an interesting symmetry that holds between the partner
attractors:
\begin{figure}[btp]
\begin{center}
  \includegraphics[width=80mm,clip]{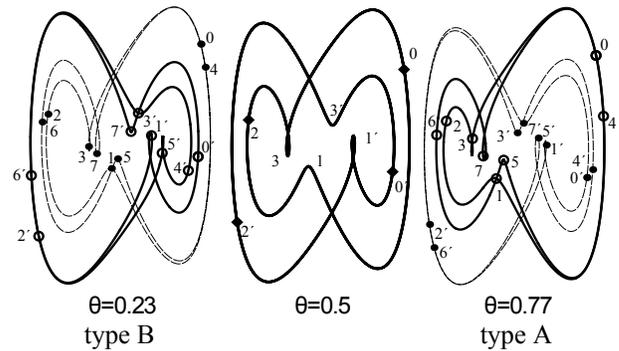}
  \caption{
  \label{Fig:fig6}
  Two-clustered attractors in GCFL with $N=100$, $k=0.241$, $\varepsilon=11\times 10^{-5}$.
  The cluster population is examined at $t=100\times 2\pi)$ to determine $\theta=N_+/(N_++N_-)$.
  The motion of majority (minority) cluster thereafter is depicted by white
  (black) circles on its orbits depicted by a bold (dashed) line at every $\pi/2$ step.
  In type A (B) attractor with $\vartheta >0.5$ $(\vartheta <0.5)$, the positive (negative) cluster is the majority. One finds an exact symmetry \eqref{cluster_pairing} and an approximate symmetry \eqref{cluster_pairing_approx}.
  }
\end{center}
\end{figure}
\begin{align}
\label{cluster_pairing}
  (T_{\pi}\otimes P)(X_{A}(t),& Y_{A}(t))^{[C]} \equiv (- X_A(t+\pi), - Y_A(t+\pi))^{(C)} \nonumber \\
  &=(X_B(t),Y_B(t))^{(C)}~~ (C=L, S),
\end{align}
where $C=L,S$ denote the large (majority) and small (minority) clusters, respectively.
This symmetry holds for majority and minority clusters separately.
Note that at the limit $\vartheta \rightarrow 0.5$, this relation reduces
to the symmetry observed in the central event. This, in turn, implies that
we also have an approximate symmetry relation
\begin{align}
\label{cluster_pairing_approx}
  ((X_{A}(t),& Y_{A}(t))^{(L,S)} \approx (X_B(t),Y_B(t))^{(S,L)},
\end{align}
as confirmed by Fig.~\ref{Fig:fig6}.

\begin{figure*}[!bpt]
\begin{center}
  \includegraphics[width=139.2mm,clip]{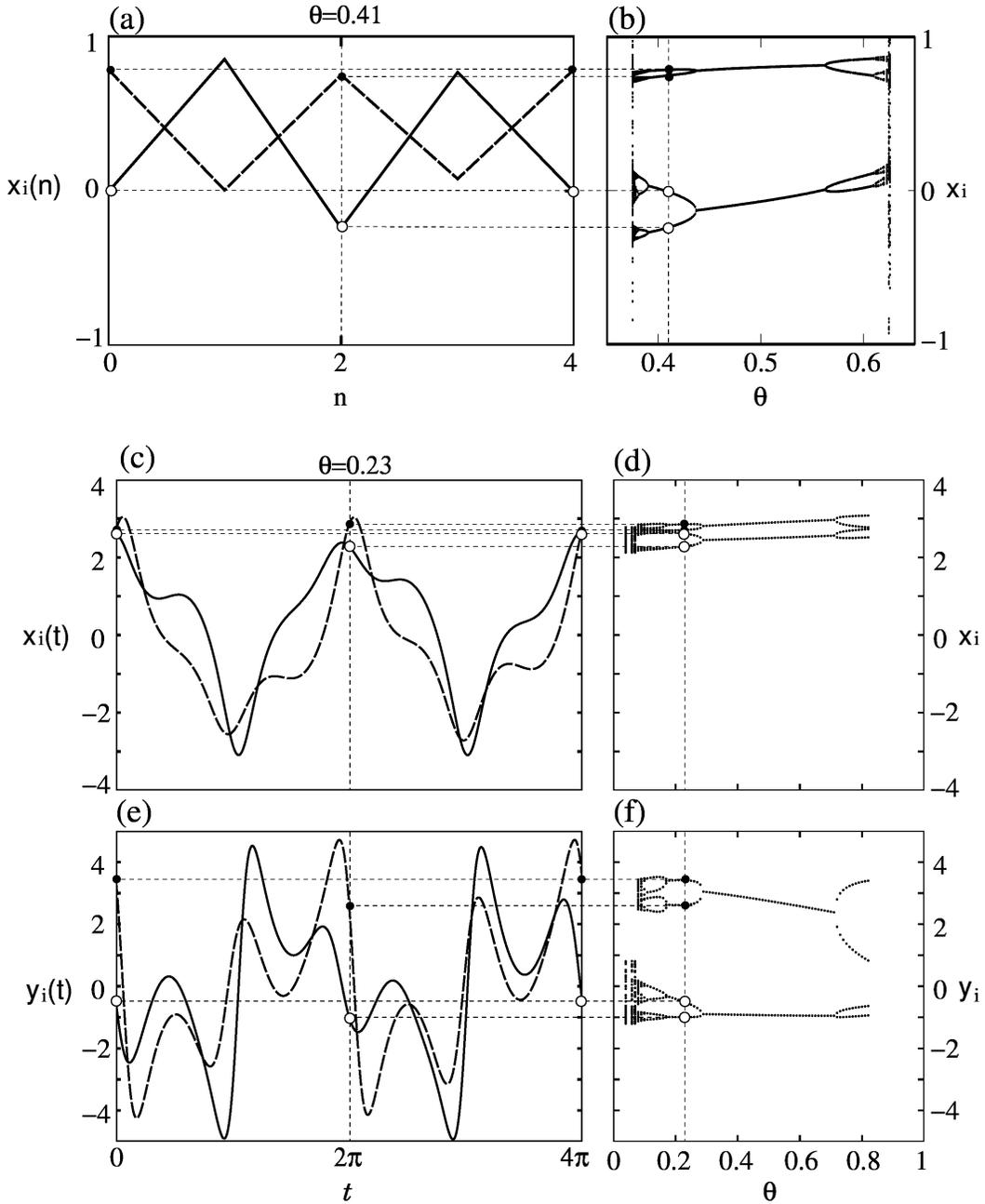}
  \caption{\label{Fig:fig7}
The $\vartheta$-bifurcation diagrams (right)
associated with the evolution plots of the sample events (left) illustrate
how diagrams are calculated.
The logistic GCML [a, b] and the Duffing GCFL x-coordinates [c,d], y-coordinates [e,f].
In the evolution plot, majority and minority cluster orbits are depicted by solid and dashed lines,
respectively.
Their positions, at every even step for GCML and at every even multiple of $\pi$
for GCFL, are depicted by white and black circles for the majority and minority clusters, respectively,
and they contribute to the $\vartheta$-bifurcation diagrams altogether four points (two for each cluster),
as indicated by horizontal fine dotted lines.
}
\end{center}
\end{figure*}

Given that the dynamics of the $N$ flows is reduced by synchronization to that of two clusters,
that is, assuming the two-clustered configuration,
it can be seen that the symmetry \eqref{pairwise-attractors} of the single Duffing oscillator
is lifted to \eqref{cluster_pairing}. We show below that
if \eqref{cluster_pairing} holds at time $t$, then it also holds
at time $t+\Delta t$.

The GCFL evolution is an iteration of the two-step process, so
let us start with the first step. For the two-clustered configuration,
this leads to
\begin{eqnarray}
   X^{mid}(t+\Delta t) &=& F(X(t), Y(t)) \equiv X(t) +Y(t) \Delta t, \nonumber\\
\label{cluster_evolution_step}
   Y^{mid}(t+\Delta t) &=& G(X(t), Y(t), t) \\
                       &\equiv& Y(t) +[-k Y(t) -X(t)^3 +A \cos t]\Delta t.\nonumber
\end{eqnarray}
This applies to both majority and minority clusters and to both type A and B attractors.
First, let us examine $Y^{mid}$, where $G$ has an explicit $t$ dependence.
We have
\begin{eqnarray*}
\label{respecting_symmetry}
   Y_A^{mid}(t+ \Delta t) &=& G(X_A(t), Y_A(t), t) \\
   &=& G(-X_B(t+\pi),-Y_B(t+\pi),t)\\
   &=& - G(X_B(t+\pi),Y_B(t+\pi),t+\pi)\\
   &=& - Y_B^{mid}(t+ \pi + \Delta t).
\end{eqnarray*}
The first and forth equalities are from
\eqref{cluster_evolution_step},
the second follows from the assumption at $t$, and the third
from the symmetry of the Duffing equation \eqref{pairwise-attractors}
used in the first step.
Thus, we obtain $Y_A^{mid}(t+ \Delta t) = - Y_B^{mid}(t+ \pi + \Delta t)$.
For $X^{mid}$, $F$ does not involve $t$ and we immediately find that
$X_A^{mid}(t+ \Delta t) = - X_B^{mid}(t+ \pi + \Delta t)$.
Thus, we have shown that \eqref{cluster_pairing} holds
for both $X^{mid}$ and $Y^{mid}$.
The next step is the interaction through the mean field. For the two-clustered configuration,
for the $X$ coordinates, we have
\begin{eqnarray*}
\label{cluster_interaction_step}
   X^{(L)}(t+ \Delta t) = (1-\varepsilon_D \vartheta_S) X^{(L,mid)} + \varepsilon_D \vartheta_S X^{(S,mid)} \\
   X^{(S)}(t+ \Delta t) = (1-\varepsilon_D \vartheta_L) X^{(S,mid)} + \varepsilon_D \vartheta_L X^{(L,mid)}
\end{eqnarray*}
and the same also holds for the $Y$ coordinates. As these are a linear mapping,
\eqref{cluster_pairing} is apparently kept.
Thus, we have shown that \eqref{cluster_pairing} is the symmetry of the GCFL evolution.

\subsubsection{A comparison between the GCML and GCFL $\vartheta$-bifurcations}
With this symmetry in mind, let us compare the $\vartheta$-bifurcation in GCML and that in GCFL in detail in Fig.~\ref{Fig:fig7}.
%%%%%
We use a representation devised by Kaneko \cite{kaneko1}.
As for GCML, we draw all values of the maps at $\vartheta$
in the final cluster attractor at all even time steps.
First, consider an event of a two-clustered attractor with $\vartheta=0.5$
 (the evolution is not drawn).
This is the basic attractor in the GCML two-cluster regime, in the sense that it is
free from $\vartheta$-bifurcation.
The two clusters oscillate mutually opposite in phase around their mean field,
and each one returns to its previous position at every two steps. Thus, the basic motion
produces two points at $\vartheta=0.5$.
In Fig.~\ref{Fig:fig7}, (a) a sample event with $\vartheta=0.41$ is exhibited.
The orbits are bifurcated to period four for high population imbalance,
and they contribute four points (two for each cluster) to the $\vartheta$-bifurcation diagram [(b)].
The omission of odd step data serves to display the $\vartheta$-bifurcation
of two clusters separately in one diagram.

As for the Duffing GCFL, the basic motion at $\vartheta=0.5$ is period one ($T=2\pi$),
as discussed in Fig.~\ref{Fig:fig6}.
Therefore, we take the Poincar\'e shot of all flows at every multiple of $2 \pi$
and plot them at the $\vartheta$ to construct the $\vartheta$-bifurcation diagram.
In Fig.~\ref{Fig:fig7}(c) and (e), we show a sample event with $\vartheta=0.23$.
Due to the population imbalance, the two-clustered attractor is bifurcated to period two ($T=4 \pi$),
and the cluster orbits again contribute four points in the $\vartheta$-bifurcation diagram.
In GCFL, the mean field $h_{x,y}(t)$ oscillates because of the external periodic force
and the two clusters mutually oscillate around this oscillating $h_{x,y}(t)$. However, at the
Poincar\'e shot, we clearly observe that the two-clustered attractor in GCFL shows
a remarkable $\vartheta$-bifurcation, which is quite similar to that in GCML.

It is worth summarizing two clear distinctions between the GCML and GCFL
$\vartheta$-bifurcation.

\noindent
(i)
The basic motion at $\vartheta=0.5$ in the two-clustered attractor regime is
period two in GCML and period one ($T=2 \pi$) in GCFL.
That is, two iteration steps in GCML correspond to one rotation in the Duffing GCFL:
\begin{align}
\label{system_correspondence}
\begin{array}{ccc}
               & \text{2-clustered regime}  &  \\
  \text{GCML}  &  \Longleftrightarrow & \text{Duffing GCFL} \\
 n\rightarrow n+2 &  & t\rightarrow t + 2 \pi .% \eqref{GCML_mapping}\ast \eqref{GCML_interaction} &  &
% \prod \left(%%\eqref{GCFL_evolution}\ast \eqref{GCFL_interaction}\right)
\end{array}
\end{align}
The $\vartheta$-bifurcation occurs successively from this basic attractor
with increasing population imbalance.

The matching of the nonlinearity between the elements
is fixed in \ref{subsec:Duffing construction}
on the basis that one iteration of a free map corresponds to
one rotation ($T=2\pi$) of a free Duffing flow.
There is no contradiction because in \eqref{system_correspondence},
the correspondence between two systems is considered,
where interaction is in action and system nonlinearities are
reduced. We discuss this point in detail in the following subsection.

\noindent
(ii)
In GCML, each of the two clusters evolves in its one-dimensional orbit
and two orbits of the clusters approximately agree with each other (precisely if $\vartheta=0.5$)
modulo a shift of one step.
In the Duffing GCFL, on the other hand, the cluster attractor is two dimensional
under the $T \otimes P_\pi$ symmetry \eqref{cluster_pairing}.
Both (i) and (ii) together realize quite similar $\vartheta$-bifurcation diagrams
for GCML and GCFL.

\begin{figure*}[!tbp]
        \begin{center}
          \includegraphics[width=140mm,clip]{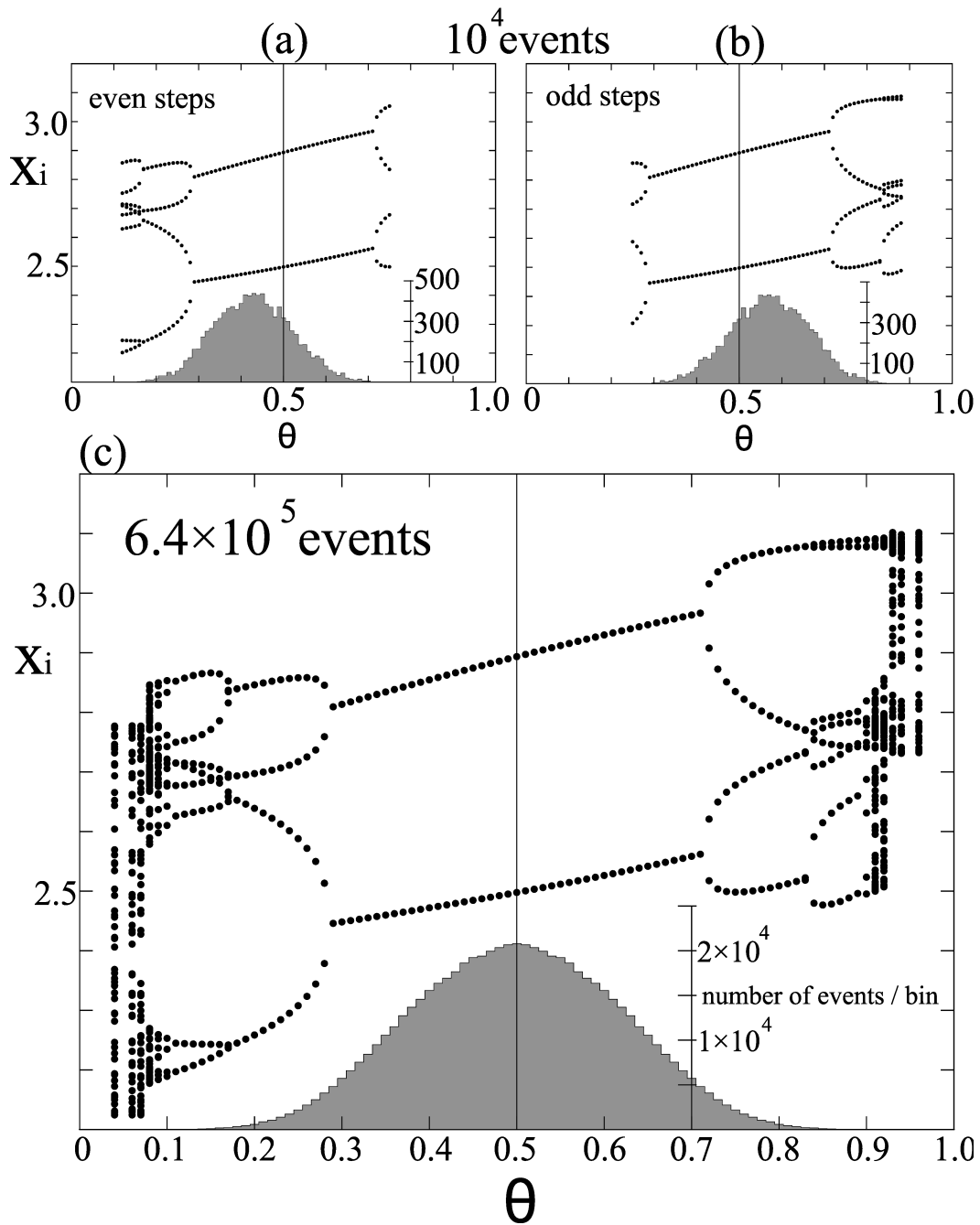}
          \caption{\label{Fig:fig8}
        GCFL $\vartheta$-bifurcation diagrams.
        (a) is the same with Fig.~\ref{Fig:fig7}(d)(flows
        at every even multiple of $2\pi$) except that the number of
        initial configurations is limited to $10^4$
        to show $\vartheta$ polarization.
        In the event formation probability below,
        one observes that type B is clearly more frequently produced.
        (b) is deduced via \eqref{symmetrytransformation} from the same $10^4$ data,
        but orbits at odd multiples of $2\pi$ are converted by the relation
        \eqref{cluster_pairing}.
        (c) is our final GCFL $\vartheta$-bifurcation tree constructed from
        a total of $6.4 \times 10^5$ events; both even and odd steps data
        are combined).}
        \end{center}
\end{figure*}

\subsubsection{Polarization of the $\vartheta$-bifurcation tree in GCFL}
 Let us add a brief remark here.
The GCFL $\vartheta$-bifurcation diagram is
strongly polarized (the larger $\vartheta$ forks are missing)
in contrast to the unpolarized GCML diagram.
This is possible since the external oscillatory
force may favor type-B rather than type-A attractors in the formation process,
while GCML is autonomous.
In order to overcome unwelcome missing bifurcation trees in the larger $\vartheta$ side,
we can use the relation \eqref{cluster_pairing}.
If we measure the type-B event (with some $\vartheta=\vartheta_0$) at every {\it odd} multiple
of $\pi$, it supplies equivalent data for the type-A event (with $\vartheta = 1-\vartheta_0)$ at
every {\it even} multiple of $\pi$.
Fig.~\ref{Fig:fig8} presents the GCFL polarization-free
$\vartheta$-bifurcation diagram thus obtained.
To our knowledge, this is the first observation of
$\vartheta$-bifurcation in the globally coupled Duffing flows
and its comparison with $\vartheta$-bifurcation in GCML.

\subsubsection{A comment on a posi-nega switch}
It would be interesting to investigate if a posi-nega switch found in the  GCML \cite{kaneko1} can also be
realized in GCFL, but we defer this for the following reasons.
In GCML, the two periodic clusters have negative Lyapunov exponents.\footnote{
Precisely, for the two-clustered attractor state with populations $N_1$ and $N_2$,
the Lyapunov exponents generally divide into two sets.
The first one consists of two exponents, $N_1-1$ and $N_2-1$ fold degenerate, respectively.
These are responsible for the contraction of the maps to the center of the respective cluster.
For the two clusters of concern, we have measured these as negative \cite{tsst-arob}.
The second consists of two exponents that are responsible for the
stability of the cluster orbits and they are also negative.}
In several steps of iteration, the variance of map coordinates
in each cluster becomes less than the machine epsilon \cite{tsst-arob}.
One could introduce low level noises to avoid numerical collisions,
but then the analysis of the posi-nega switch will involve technical subtleties.
The GCFL attractors have also negative Lyapunov exponents and share the same
difficulty in the numerical analysis. Furthermore,
we have checked that the GCFL two-clustered attractors are robust
in the sense that the threshold value $\vartheta_{th}$ to the decay into
a chaotic transient process is quite high ($\vartheta_{th}\approx 0.9$)
(or quite low $\vartheta_{th}\approx 0.1$).
This is a negative indication for an implementation of the switch,
because, in order to enhance the population imbalance beyond the threshold,
many elements must be transferred from the minority to the majority cluster
and, then, the above collision problem becomes more severe in the GCFL than in the GCML.
\section{TWO-CLUSTERED DYNAMICS OF THE DUFFING GCFL}
\subsection{Nonlinearity reduction}
The periodic attractor in the two-clustered regime is a consequence of the reduction
in the high nonlinearity of the elements caused by the averaging interaction via the mean field.
We first recapitulate this issue in GCML,
where some quantitative understanding is possible
on the basis of the Perez-Cerdeira transformation for the quadratic map.
We then numerically investigate the case of GCFL in some detail.

\subsubsection{Case of GCML;
foliation curves of periodic window dynamics of element maps
into the (a, $\varepsilon$) plane}
\label{subsec:GCML_foliation}

%%%%%%%%%%%%%%%%%%%%%%%%%%%%
%%%%%%%%%%%%%%%%%%%%%%%%%%%%
%%%%%%%%%%%% Fig7  %%%%%%%%%
%%%%%%%%%%%%%%%%%%%%%%%%%%%%
%%%%%%%%%%%%%%%%%%%%%%%%%%%%
\begin{figure*}[!tbp]
  \begin{center}
     \includegraphics[width=165mm,height=140mm,clip]{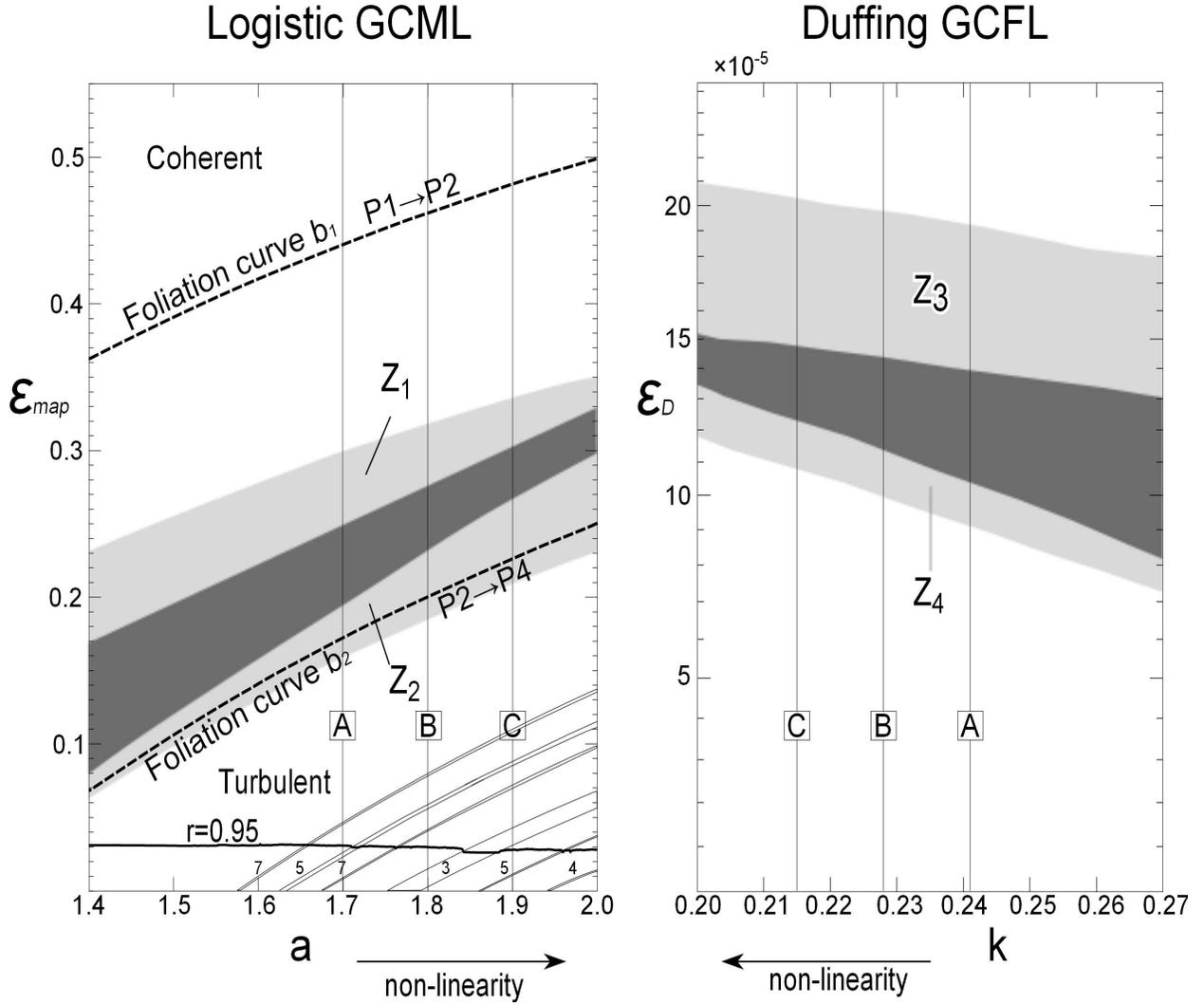}
     \caption{\label{Fig:fig9}
        The band of the two-clustered dynamics of GCML (left) and GCFL (right)
        in the parameter spaces ($a,\varepsilon_\text{map}$) and ($k,\varepsilon_D$), respectively.
        (For comparison, the $\varepsilon_D$-axis is adjusted to
        the $\varepsilon_\text{map}$-axis using the relation \eqref{epsilon_map_flow_correspondence}.)
        In the black band, the formation probability of the two-clustered periodic attractor is more than 90%,
        while in the adjacent gray regions, the basin volume is shared more by nearby dynamics
        (coherent chaos in $Z_1$ and $Z_3$, multiclustered attractors in $Z_2$,$C \otimes C$ in $Z_4$),
        and toward the outer edges of the gray bands, the partition of the two-clustered periodic
        attractor decreases down to 10\% (Fig.~\ref{Fig:fig5}).
        The judgment of final states is performed by inspection after $10^3$ iteration steps from
        an initial random start. If it is performed after $10^2$ steps, $Z_1$ is partially ordered.
         For GCML, the foliation curves of the dominant windows are also shown; in particular, $b_1$ and $b_2$, respectively, emanate
        from the first and second bifurcation points ($p1\rightarrow p2$ and $p2\rightarrow p4$) (ref.~\cite{tsst-arob,tskk-pre}).
        }
  \end{center}
\end{figure*}

In GCML, all element maps have high nonlinearity, and
the maps would evolve independently in random motion if the interaction would be
switched off. However, at every iteration step, the map coordinates
are uniformly contracted to the mean field and coherence
is introduced. The nonlinearity of the system is effectively reduced
and the periodic clustered attractors are formed.

Now, interestingly, there is a case in which this reduction can be
quantitatively estimated.
This is the case of the maximally symmetric cluster attractors (MSCAs)
formed in the turbulent regime. We refer to \cite{tskk-pre} for a detailed account.
In the MSCA, the maps synchronize in almost equally populated $c$ clusters
that oscillate in period $p=c$ around their mean field.
The most prominent one is the $p3c3$ MSCA. In such an MSCA,
the mean field is almost constant because of the population symmetry,
and hence, it is possible to estimate to what extent the nonlinearity of the element maps
is reduced. In fact, as the mean field is approximately a
constant $h_n \approx h^{*}$,
the GCML evolution equation leads to
\begin{equation}
\label{GCML_evolution_const_h}
x_{i}(n+1) =(1-\varepsilon)\left[1-a (x_i(n))^2 \right]+\varepsilon h^*,~~i=1,\cdots,N.
\end{equation}
Then, the scaled maps defined by the Perez-Cerdeira transformation \cite{perezcerdeira}
\begin{equation}
\label{perezcerdeira}
    y_i(n) \equiv \frac{x_i(n)}{1 - \varepsilon +\varepsilon h^*},
\end{equation}
all equally follow the same scaled map
\begin{align}
    y_{i} (n + 1) &= 1 - b\left(x_i (n)\right)^2,~~i=1,\cdots,N.
\label{reducedmap}
\end{align}
The nonlinearity is reduced from $a$ to
    $b = r a$,
where
\begin{align}
\label{reduction-factor}
   r = (1 - \varepsilon)\left[1 - \varepsilon (1 - h^*)\right] < 1
\end{align}
is the rate of nonlinearity reduction.
A good rule of thumb is $r \approx 1-2\varepsilon$.

We can go a step further.
Taking the average of \eqref{perezcerdeira} over the elements, we obtain
\begin{equation}
\label{perezcerdeira2}
 \frac{h^{*}}{1 - \varepsilon +\varepsilon h^*} = \langle y \rangle =  y^*_b
 \end{equation},
 where $y^*_b$ is the long-time average of a single map $y_b(n)$ with nonlinearity $b$.
\footnote{
$\langle y \rangle$ is, for the MSCA, an equal-weight average
of the cluster orbits over $p=c$ clusters. But, again
for the MSCA, this is nothing but the average of the cluster orbit points $Y_I(n)$ over period $p$.
This, in turn, is also the case with $y^*_b$. Namely,
$
 \langle y \rangle
=\sum_{I=1}^c Y_{I}(n) (\forall n)
=\sum_{n=n_0}^{n_0+p-1} Y_{I}(n) (\forall I)
= y^*_b \equiv \lim_{T\rightarrow \infty}
 \frac{1}{T} \sum_{n=T_0}^{T_0+T} y_{b^*}(n)
$
\cite{tskk-pre}.
}
Solving \eqref{reduction-factor} and \eqref{perezcerdeira2}
for $a$ and $\varepsilon$ at a given $b$ and $r$,
one obtains the parametric form of a curve (labeled by $b$ and develops
with $r$) that describes the foliation of dynamics of a period $p$ window
into the $(a,\varepsilon)$ plane.
\footnote{
Our formulation of the foliation curves \cite{tskk-pre} was inspired
by \cite{just}.}
\begin{align}
\label{foliation_curves}
 a^{(b)}(r)&= \frac{b}{r},\\
  \varepsilon^{(b)}(r) &=  1 - \frac{r y_b^*}{2}
-\sqrt{r(1- y_b^*) + \left(\frac{ry_b^*}{2}\right)^2}
\end{align}
We generally observe the period $p$ MSCA ($r\gtrsim 0.95$) and its remnant state ($r\lesssim 0.95$)
in the $(a,\varepsilon)$ region, which lie between the two foliation curves emanating from
$b_1$ and $b_2$ (indicating nonlinearity at the $p$ window's opening and at the first bifurcation
in the $p$ window, respectively). \footnote{
The most prominent $p3c3$ is an exception. It is observed down to $r \sim 0.84$.}
In the adjacent region (with larger $\varepsilon$), period $p$, $c$-clustered states (with
$p>c$) are formed.
The foliation curves are exhibited in Fig.~\ref{Fig:fig9}-GCML.

%--
We should now clarify the logic.
The foliation curves are derived assuming the existence of the MSCAs.
If a $p=c$ MSCA is formed, it must be within the $(a,\varepsilon)$ region
dictated by the foliation curves of the period $p$ window.
Furthermore, we can prove that at the MSCA, all the Lyapunov exponents are negative
and the MSCA is a stable attractor \cite{tskk-pre}. However, the converse is not necessarily true.
There is no guarantee that the MSCA is formed everywhere in the range specified
by the foliation curves; indeed for $r \lesssim 0.95$, we observe only the remnants of
MSCAs ($p>c$ clustered attractors) as the valleys (peaks) in the distribution of the
mean squared deviation of the mean field $h(n)$ \cite{tskk-pre,tskktm-arob09}.

Now, in the two-clustered regime of concern, the maps divide themselves into two
clusters, and the two clusters oscillate mutually around the mean field.
Here the attractor is robust in that it allows the population
imbalance accompanied by the attractor bifurcation. However,
in the particular case of $\vartheta=0.5$, we can regard it just as a
$p=c=2$ MSCA. With this concept, we can apply the above formulation
to predict the possible region in which the two-clustered attractor may be formed.

In Table.~\ref{tableGCMLfoliation} we compare
the observed $\varepsilon$ range of the two-clustered regime
with the predicted bounds from
the foliation of the period-two window dynamics.
For the latter, the two curves from $b_1=0.75$ and $b_2=1.25$ are used,
which are the opening point of the $p2$ window
(the threshold of $p1\rightarrow p2$) and the threshold of $p2 \rightarrow p4$, respectively.
Table.~\ref{tableGCMLfoliation} shows that the observed range ($\varepsilon_2,\varepsilon_3$)
is inside the predicted bounds.
Moreover, Fig.~\ref{Fig:fig9}-GCML shows that the observed band for the formation of two-clustered
attractors runs just in the center of the above two foliation curves.

%------
%\input table1.tex
\begingroup
% Table I
\squeezetable
\begin{table}[hbp]
        \caption{\label{tableGCMLfoliation}
        Two-clustered regime in GCML\\}
        \begin{ruledtabular}
        \begin{tabular}{lcr}
        \textrm{Nonlinearity\footnote{Matching points A, B, and C
        with the corresponding nonlinear parameter $a$ in the parenthesis
        (see Fig.~\ref{Fig:fig1}(a)).}
        }
        &
        \textrm{observed $\varepsilon$-range\footnote{
        Period two, two-clustered attractor ($p2c2$) formation occupies more than
        90\% of the basin volume in $\varepsilon_2-\varepsilon_3$, and the fraction decreases
        down to 10\% toward $\varepsilon_1$ and $\varepsilon_4$.
        The rest of the basin volume is occupied by the formation of three or four clusters in random
        motion in $\varepsilon_1-\varepsilon_2$, and single-clustered chaos
        in $\varepsilon_3-\varepsilon_4$, respectively.}
        %\TM
        }
        &
        \textrm{Prediction(bounds)\footnote{The prediction for the bounds based on $p2c2$ foliation curves.
        The upper (lower) bound
        from $p1\rightarrow p2$ ($p2\rightarrow p4$) bifurcation threshold of the element map.
        }        }
        \\
	 & $\varepsilon_{1}$ - $\varepsilon_{2}$ - $\varepsilon_{3}$ - $\varepsilon_{4}$ & \\
        \colrule
        A (a=1.70) & 0.16 - 0.19 - 0.25 - 0.30 & 0.17 - 0.44 \\
	 B (a=1.80) & 0.18 - 0.23 - 0.28 - 0.32 & 0.20 - 0.46 \\
        C (a=1.90) & 0.21 - 0.27 - 0.30 - 0.34 & 0.23 - 0.48 \\
        \end{tabular}
        \end{ruledtabular}
\end{table}

\endgroup

%------
%------

\subsubsection{Case of the Duffing GCFL}
\label{reductionGCFL}

Comparing Fig.~\ref{Fig:fig4} and Fig.~\ref{Fig:fig5},
one clearly observes that, as is the case for GCML, the Duffing GCFL reduces the system
nonlinearity caused by the averaging interaction via the mean field.
The $\varepsilon$ ranges for the two-clustered regime
at A,B, and C are listed in Table.~\ref{tableGCFLfoliation}.
Again, with increasing nonlinearity,
the $\varepsilon$ necessary to maintain the two-clustered attractor
shifts toward the higher end.

%------
%\input table2.tex
\begingroup
% Table II
\squeezetable
\begin{table}[!thb]
        \caption{
        \label{tableGCFLfoliation}
        The two-clustered regime in The DuffingGCFL\\}
        \begin{ruledtabular}
        \begin{tabular}{lcr}
        \textrm{Nonlinearity\footnote{The matching points A, B, and C
        with the corresponding friction parameter $k$ in the parenthesis
        (Fig.~\ref{Fig:fig1}(b)).}} &
        \textrm{observed $\varepsilon_D$-range\footnote{$\varepsilon_D$ in unit of $10^{-5}$ when $\Delta t = 10^{-3}$.
        Period one, two-clustered attractor ($p1c2$) formation occupies more than
        90\% of the basin volume in $\varepsilon_2-\varepsilon_3$, which decreases
        down to 10\% toward $\varepsilon_1$ and $\varepsilon_4$.
        In $\varepsilon_1-\varepsilon_2$, the rest of the basin volume is occupied by two clusters, each in random motion, while in $\varepsilon_3-\varepsilon_4$, by
        a single-clustered chaotic attractor whose topology is chiral for point A and self-dual
        for points B and C (see insets in Fig.~\ref{Fig:fig5}).}
        }
        &
        \textrm{Estimation from GCMLL\footnote{Estimation for
        $\varepsilon_D$-range via \eqref{epsilon_map_flow_correspondence} from $\varepsilon_\text{map}$-range in Table \ref{tableGCMLfoliation}.}}\\
         & $\varepsilon_1$ ~-~  $\varepsilon_2$  ~-~ $\varepsilon_3$  ~-~ $\varepsilon_4$ & \\
        \colrule
        A ($k=0.241$) & 9.3 - 10.4 - 13.4 - 19.0  & 5.5 - 6.9 - $~$ 9.1 - 11.3 \\
        B ($k=0.228$) & 10.1 - 11.4 - 14.0 - 19.5 & 6.5 - 8.4 - 10.2 - 12.2 \\
        C ($k=0.215$) & 10.8 - 12.4 - 14.3 - 20.1 & 7.4 - 9.9 - 11.5 - 13.0 \\
        \end{tabular}
        \end{ruledtabular}
\end{table}
\vspace{5mm}
\endgroup
%------
%------
Let us examine how the observed parameter range in GCFL compares with that in GCML.

Since two steps in GCML correspond to one cycle ($2\pi$) in GCFL (see \eqref{system_correspondence}),
one may roughly predict the range of $\varepsilon$ in GCFL
via the data of $\varepsilon$ in GCFL by a relation
\begin{equation}
\label{epsilon_map_flow_correspondence}
 (1-\varepsilon^\text{estimation}_{\mathrm{D}})^{{{2 \pi}}/{\Delta t}} \approx \left(1-\varepsilon_{\rm map}\right)^2,
\end{equation}
each side expressing the contraction rate in respective models.
%%%%%%%%%%%%%%%%%%%%%%
\footnote{
We are aware that this is only a crude approximation
because all the $2\pi/\Delta t$ of the one-step
contraction ($1-\varepsilon_D$) factors are gathered and multiplied
into a single factor after (unjustifiably) commuting them through
all the small time ($\Delta t$) evolution steps.
However, we have checked, varying $\Delta t$ from $10^{-3}$ down to $10^{-5}$,
that numerically measured $\varepsilon_D$
keeps the left hand side of \eqref{epsilon_map_flow_correspondence} almost
invariant. This supports the above approximation to some extent.}
%%%%%%%%%%%%%%%%%%%%%%
 Note that the matching of $k$ and $a$ is considered between
a free flow and a free map and then
one period ($2\pi$) of a flow naturally corresponds to
one iteration step of a map.
On the other hand, \eqref{epsilon_map_flow_correspondence}
is considered between flows and maps in actual interaction,
realizing the correspondence \eqref{system_correspondence}.
%%%%%%%%%%%%%%%%%%%%%%
Table~\ref{tableGCFLfoliation} shows that the prediction
for the two-clustered attractor range roughly agrees with the observed $\varepsilon_D$ range,
although generally, the latter turns out to be somewhat higher than the former simple prediction.
%%%%%%%%%%%%%%%%%%%%%%
%%%%%%%%%%%%%%%%%%%%%%

To analyze this issue globally in quantitative terms,
we show in the right panel of Fig.~\ref{Fig:fig9},
the band of two-clustered dynamics on the $(k, \varepsilon_D)$ plane,
where the $\varepsilon_D$ axis is adjusted using the relation \eqref{epsilon_map_flow_correspondence} to facilitate the comparison with the
GCML case in the left panel.
We find that the two-cluster band in the Duffing GCFL
runs in the $(k,\varepsilon_D)$ plane
roughly like the foliation band of $p2c2$ attractor in GCML
in the $(a,\varepsilon_\text{map})$ plane but with two notable
differences: \\
(1) the GCFL band turns out with a somewhat higher coupling,\\
(2)the GCFL band is less tilted than the GCML one.
These differences can be understood as follows.
For simplicity, let us consider the basic two-clustered
attractor with $\vartheta=0.5$, which is free from $\vartheta$-bifurcation.
In GCML, this is period two ($p2c2$), while in the Duffing GCFL, it is period one,
as discussed in \ref{subsec:theta bifurcation} (see especially
\eqref{system_correspondence}).
Pairwise clusters in GCFL {\it (each period one)} emulate the GCML $p2c2$ attractor.
Therefore, while the nonlinearity is reduced from chaos
to period two in GCML, it should be reduced deeply down to
period one in GCFL, which requires higher coupling at the equivalent nonlinearity.
%%%%%%

%%%%%%%%%%%%%%%%%%%%%%%%%%%%
%%%%%%%%%%%%%%%%%%%%%%%%%%%%
%%%%%%%%%%%  Fig7b %%%%%%%%%
%%%%%%%%%%%%%%%%%%%%%%%%%%%%
%%%%%%%%%%%%%%%%%%%%%%%%%%%%
\begin{figure*}[!bht]
\begin{center}
  \includegraphics[width=134mm,clip]{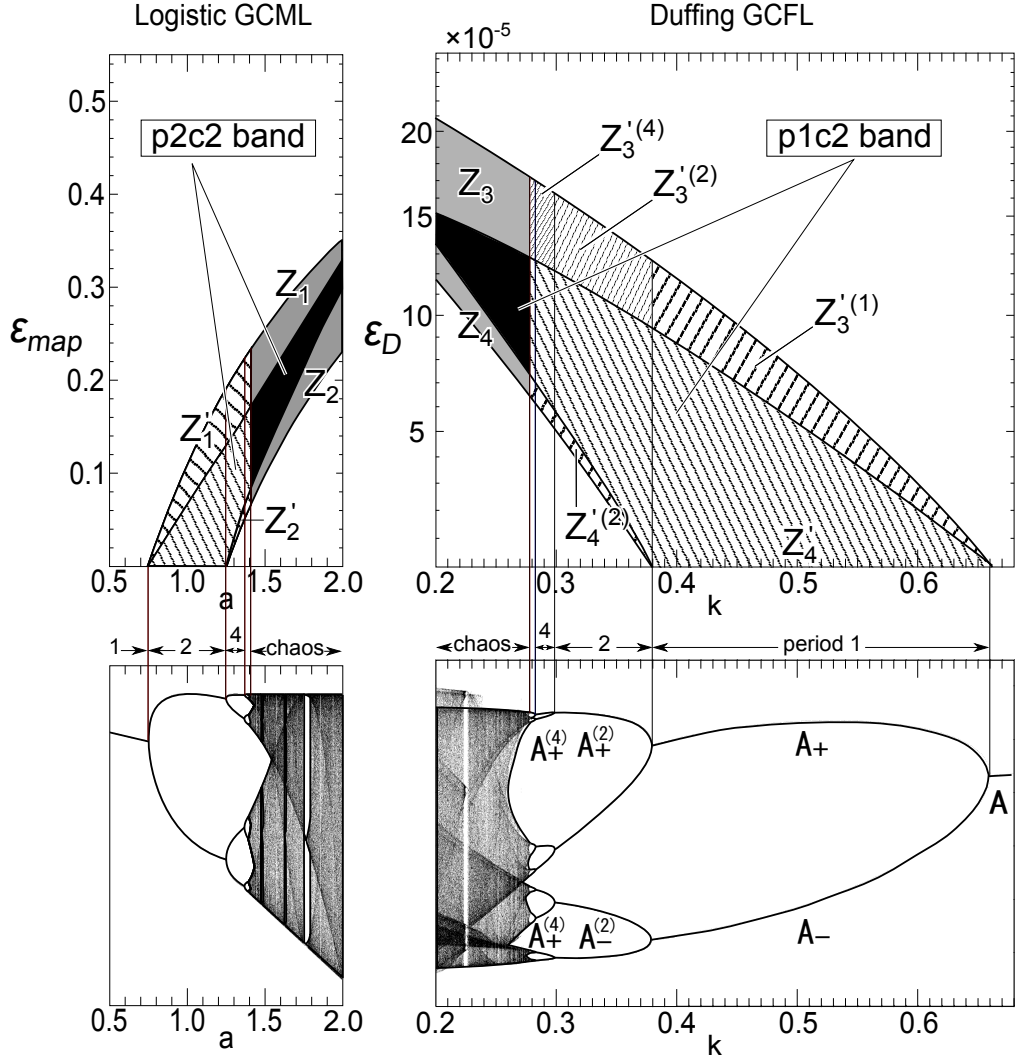}
\caption{\label{Fig:fig10}
The upper diagram exhibits the extended band of the two-clustered dynamics and the lower
diagram exhibits full bifurcation trees of the elements [GCML (left) and GCFL (right)].
The format for the bands is the same as that in Fig.~\ref{Fig:fig9}
except that the extended regions where the elements are periodic
are depicted by hatched lines.
While Fig.~\ref{Fig:fig9} exhibits
the nonlinearity reduction of the chaotic elements to the periodic two-clustered attractor,
the extension in this figure exhibits that of periodic elements to
the periodic two-clustered attractor.
The compositions are as follows.
GCML: two-clustered attractor with period two ($p2c2$) in the central band,
$p2c1$ in $Z_1^\prime$, $p4c4$ in $Z_2^\prime$. GCFL: two-clustered attractor with period one ($p1c2$) in the central band, $p4c1,p2c1,p1c1$ in $Z_3^{\prime(4),(2),(1)}$, respectively, and  $p2c2$ in $Z_4^{\prime}$.
The phases in the extended region can be easily understood from
the bottom bifurcation diagram for the elements. For a single Duffing flow with $A=7.5$,
the periodic region is $k>0.277$,
the period-two attractor ($A^{(2)}_{+}$, $A^{(2)}_{-}$) in $k \in (0.299,0.38)$,
and the period-one attractor ($A_{+}$, $A_{-}$) in $k \in (0.38,0.66)$.
}
\end{center}
\end{figure*}

%%%%%%
To quantitatively verify this explanation, we have
extended the range of $k$ ($a$) (Fig.~\ref{Fig:fig10}) such that it covers
the periodic regime of the element flow (map).
In the extended region, the elements are already periodic and hence, if the coupling is
sufficiently strong to maintain overall coherence, a single clustered attractor
is formed, which evolves with the same periodicity as that of the element
($Z_1^\prime$ and $Z_3^{\prime(4),(2),(1)}$ for GCML and GCFL, respectively.)
If, on the other hand, the coupling becomes sufficiently small, a multiclustered periodic
attractor is formed in GCML ($Z_2^\prime$), while in GCFL,
the periodic external force compels the flows to synchronize in two clusters,
one subject to $A_{+}^{(2)}$($A_{+}$) and the other to $A_{-}^{(2)}$($A_{-}$)
depending on the bifurcation state of the element flows ($Z_4^{\prime(2)}$, $Z_4^\prime$).
In the intermediate coupling, the extension of the central band of Fig.~\ref{Fig:fig9}
is formed for both GCML and GCFL. We clearly observe that the two-cluster band of  GCFL is
smoothly connected to the period-one window of the element
flow at $\varepsilon \rightarrow 0$, just as the GCML two-clustered band
flows along foliation curves to the period-two window.

\section{Conclusion}

 We have constructed a Duffing GCFL as a natural extension of the logistic GCML
and investigated to what extent the intriguing features of the logistic GCML
are inherited by our model. This, in turn, amounts to an explanation of the extended
universality at the level of the complex systems.
To compare GCFL with GCML,
we have worked in a scheme where the element flow is adjusted to have
the same nonlinearity as that of the element map, in the sense that
their locations in the (universal) bifurcation trees match each other,
while the coupling $\varepsilon_D$ is varied as a free parameter
to seek if GCFL realizes the corresponding phenomena in GCML.
As a byproduct, we have observed that the bifurcation-tree matching leads
to remarkable agreement between the Lyapunov exponents of the elements
over the entire nonlinearity parameter value range.

 The phase diagram presented in Fig.~\ref{Fig:fig5}
clearly exhibits that in the Duffing GCFL, the same phases as those in the logistic GCML are realized
in the same order with increasing $\varepsilon$. Furthermore, the extensive analysis of the Duffing
two-clustered dynamics shows that the Duffing GCFL has the
ability of $\vartheta$-bifurcation just as GCML does (Fig.~\ref{Fig:fig8}).
We believe that this is the first systematic observation of $\vartheta$-bifurcation
in coupled flow systems.
This clearly shows that the universality between a map
and a flow is basically extended to the level of universality between the systems.

 However, we have found that there is an important distinction between the two models that
comes from the fact that a Duffing flow admits a pair of possible attractors under the symmetry
\eqref{pairwise-attractors}. As a consequence, the GCFL attractor in the two-clustered regime
consists of pairwise clusters (Fig.~\ref{Fig:fig6})
under a symmetry relation \eqref{cluster_pairing}
and each in period one (returns the same phase-space point at every $2 \pi$, modulo $\vartheta$-bifurcation). In the logistic GCML, on the other hand, the element map
has a unique attractor and the two-clustered attractor consists of two clusters, each evolving
in period two but mutually shifted by one step (i.e., in opposite phase). This distinction depicted
in Fig.~\ref{Fig:fig7} means that the nonlinearity reduction in  GCML is only down to period two,
while in GCFL, it is reduced deeply down to period one. To verify this, we have numerically
measured (extending the element nonlinearity from chaotic to periodic regime)
how the band of the two-clustered dynamics in GCFL runs in the ($k-\varepsilon$) space.
It is clearly observed (Fig.~\ref{Fig:fig10}) that the GCFL band flows into the period-one window,
just as the GCML band flows into the period-two window.
 To summarize, the Duffing GCFL basically shares the same phases and the same $\vartheta-$bifurcation
property with the logistic GCML; however, detailed examination has revealed that there is a distinction because of the difference in the basin structures of the elements. Finally, we are now investigating a GCFL of three-dimensional flows in detail.\footnote{See a brief report of preliminary results in Sec.\ref{sec:introduction}.}

\section*{ACKNOWLEDGMENTS}

TS thanks his former student, Hayato Fujigaki,
for his enthusiasm for this work.
He and TS jointly observed the population bifurcation in GCFL \cite{fs}.
It has taken almost 10 years to clarify
the Duffing GCFL attractor structures resulting from the symmetry
in the Duffing equation.


\begin{thebibliography}{9}

\bibitem{kaneko1}
K. Kaneko, Phys. Rev. Lett. {\bf 63}, 219 (1989).

\bibitem{kaneko2}
K. Kaneko, Phys. Rev. Lett. {\bf 65}, 1391 (1990).

\bibitem{kaneko3}
K. Kaneko, Physica  {\bf D41}, 137 (1990).

\bibitem{ueda}
Y. Ueda, Steady Motions Exhibited by Duffing's Equation : A Picture Book of Regular and Chaotic Motions,
National Institute for Fusion Science, Research report {\bf 434}, 1-12, 1980.
See eq.(1) and the Duffing phase diagram Fig. 1. These are also accessible in
{\it The road to chaos}, Aerial Press (1992); see eq. (17) in page 208 and Fig. 7 in page 210.

\bibitem{feigenbaumreport}
M. J. Feigenbaum, Los Alamos Science {\bf 1}, 4 (1980).

\bibitem{cvitanovic}
P. Cvitanovi\'{c} in {\it Universality in Chaos},
A beautiful account with fine figures is given in \cite{cvitanovic}.
2nd edition, Institute of Physics Publishing, Bristol and Philadelphia
(1996).

\bibitem{feigenbaumtheory}
M. J. Feigenbaum, J. Stat. Phys. {\bf 21}, 669 (1979).

%\bibitem{poincare-bendixson}
%Poincar\'e, H,  "Sur les courbes de'finies par une e'quation diffe'rentielle", Oeuvres, 1, Paris, (1892). \\
%Bendixson, Ivar, "Sur les courbes de'finies par des e'quations diffe'rentielles", Acta Mathematica (Springer Netherlands) %24 (1): 1--88, (1901).

\bibitem{lorenz} E. N. Lorenz, Journal of the Atmospheric Sciences {\bf 20} 130 (1963).

\bibitem{sparrow}
C. Sparrow, {\it The Lorenz Equations: Bifurcations, Chaos, and Strange Attractors}, New York, Springer-Verlag (1982). See page 99.

\bibitem{jackson}
E. Atlee Jackson, {\it Perspectives of nonlinear dynamics}, {\bf 1}, Cambridge University Press (1989), Fig. 7.54.
See also page 99 in \cite{sparrow}.

\bibitem{tskktm-arob08} T. Moriya, T. Shimada, H. Fujigaki, Artificial Life and Robotics, {\bf 13}, 214 (2008).
%\bibitem{smf-arob} T. Moriya, T. Shimada and H. Fujigaki, Artificial Life and Robotics, {\bf 13}, 214 (2008).

\bibitem{pikovsky}
M.G. Rosenblum, A. S. Pikovsky, and J. Kurths, Phys. Rev. Lett. {\bf 76}, 1804 (1996).
%A. S. Pikovsky, and J. Kurths, Phys. Rev. Lett. {\bf 72}, 1644 (1994).

\bibitem{pikovskybook}
%Arkady Pikovsky, Michael Rosenblum, Juergen Kurths
A. Pikovsky, M. Rosenblum, J. Kurths,
{\it Synchronization: a universal concept in nonlinear sciences},
Cambridge, 2002.

\bibitem{pecora}
L. M. Pecora and T. L. Carroll, Phys. Rev. Lett. 64, 821 (1990).\\
T. L. Carroll and L. M. Pecora, Physica 67, 126 (1993).

\bibitem{kowalski}
J. M. Kowalski, G. L. Albert, and G. W. Gross,
Phys. Rev. {\bf A42}, 6260, 1990.

\bibitem{fns}
H. Fujigaki, M. Nishi, and T. Shimada, Phys. Rev. {\bf E53}, 3192 (1996);
H. Fujigaki and T. Shimada, Phys. Rev. {\bf E55}, 2426 (1997).

\bibitem{tskk-pre} Tokuzo Shimada and Kengo Kikuchi, Phys. Rev. E{\bf 63}, 3489
    (2000).

\bibitem{tskktm-arob09} T. Shimada, K. Kubo, T. Moriya, Artificial Life and Robotics, {\bf 14}, 562 (2009).


\bibitem{shibatakaneko} T. Shibata and K. Kaneko, Physica D {\bf 124}, 177 (1998).
%Tongue-like bifurcation structures of the mean-field dynamics in a network of chaotic elements  Original Research Article
%Physica D: Nonlinear Phenomena, Volume 124, Issues 1-3, 1 December 1998, Pages 177-200
%Tatsuo Shibata, Kunihiko Kaneko

\bibitem{tsst-arob} T. Shimada and S. Tsukada,
Proc. of the Sixth Int. Symp. on Artificial Life and Robotics (AROB 6) {\bf 1} 242, 2001.

%\bibitem{tskk-arob} T. Shimada and K. Kikuchi,
%Proc. of the Sixth Int. Symp. on Artificial Life and Robotics (AROB 6) {\bf 2} 313, %2001.

\bibitem{just} W. Just, J. Stat. Phys. {\bf 79}, 429, 1995.
    7492 (1992).

\bibitem{fs}
 H. Fujigaki and T. Shimada, unpublished, 1997;
 H. Fujigaki, M. Sc. thesis, School of Science and Technology, Meiji University, Kanagawa, Japan.

%-------------------------------------------------------------

%\bibitem{tsstptp} T. Shimada and S. Tsukada, Progress of Theoretical Physics {\bf 108}, 25 (2002).

\bibitem{perezcerdeira}
G. Perez and H. A. Cerdeira, Phys. Rev. {\bf A46}, 7492 (1992).


\end{thebibliography}
\end{document}